\documentclass[a4paper]{article}

\usepackage{INTERSPEECH2020}
\usepackage{amsfonts}
\usepackage{hyperref}
\usepackage{bm}
\usepackage{float}
\usepackage{nicefrac}
\usepackage{graphicx}
\usepackage{caption}
\usepackage{subcaption}
\usepackage{multirow}
\usepackage{multicol}
\usepackage{adjustbox}
\usepackage{booktabs}
\usepackage{tablefootnote}
\usepackage{xspace}
\usepackage{algorithm}
\usepackage[noend]{algpseudocode}

\def\espresso{\textsc{Espresso}\xspace}
\def\pychain{\textsc{PyChain}\xspace}

\title{Wake Word Detection with Alignment-Free Lattice-Free MMI}
\name{Yiming Wang$^1$, Hang Lv$^{4,1}$, Daniel Povey$^{3}$, Lei Xie$^{4}$, Sanjeev Khudanpur$^{1,2}$\thanks{This work was partially supported by unrestricted gifts from \href{https://www.mobvoi.com/us}{Mobvoi Information Technology Company Limited}, and \href{https://www.apptek.com/}{Applications Technology (AppTek)}. The authors thank Hainan Xu and Tongfei Chen for valuable comments.}}
\address{
  $^1$Center for Language and Speech Processing,
  $^2$Human Language Technology Center of Excellence,\\
  Johns Hopkins University, Baltimore, MD, USA \\
  $^3$Xiaomi Inc., Beijing, China \\
  $^4$ASLP@NPU, School of Computer Science, Northwestern Polytechnical University, Xi'an, China
}
\email{\{yiming.wang,khudanpur\}@jhu.edu, \{hanglv,lxie\}@nwpu-aslp.org, dpovey@gmail.com}

\begin{document}

\maketitle
\begin{abstract}
Always-on spoken language interfaces, e.g. personal digital assistants, rely on a \emph{wake word} to start processing spoken input. We present novel methods to train a hybrid DNN/HMM wake word detection system from partially labeled training data, and to use it in on-line applications: (i) we remove the prerequisite of frame-level alignments in the LF-MMI training algorithm, permitting the use of un-transcribed training examples that are annotated only for the presence/absence of the wake word; (ii) we show that the classical  keyword/filler model must be supplemented with an explicit non-speech (silence) model for good performance; (iii) we present an FST-based decoder to perform online detection. We evaluate our methods on two real data sets, showing 50\%--90\% reduction in false rejection rates at pre-specified false alarm rates over the best previously published figures, and re-validate them on a third (large) data set.
\end{abstract}
\noindent\textbf{Index Terms}: wake word detection, lattice-free MMI, alignment free

\section{Introduction}
Wake word detection is the task of detecting a predefined keyword from a continuous stream of audio. It has become an important component in today's voice-controlled digital assistants and smart phones. Voice-controlled devices, with wake word detection system running in the background, require a low power solution. When people wish to to interact with such devices by voice, they \emph{wake up} the device by saying a predefined word like ``Alexa'' for Amazon Echo or ``Okay Google'' for Google Home. If the word is identified and accepted, the device turns on, i.e. goes into a state with higher power consumption to recognize and understand more complex spoken instructions \cite{wang2019end}.

\textbf{HMM-based keyword-filler models} are used to represent both the keyword and filler (background) models \cite{rohlicek1989continuous,rose1990hidden,szoke2005phoneme}. The keyword model consists of all valid phone sequences from the keyword, and the filler model includes all other speech and non-speech. During the decoding phase, usually the ratio of the scores with keyword graph and to the filler graph is computed for determining the presence of the wake word. With recent advances in deep learning, HMM-DNN hybrid wake word systems replace GMM-based acoustic models with a neural network to classify individual frames \cite{panchapagesan2016multi,sun2016compressed,wu2018monophone}. While the filler model for background speech is specified as an ergodic topology between speech and non-speech in \cite{panchapagesan2016multi,sun2016compressed}, it is represented as an all-phones loop in \cite{wu2018monophone}, increasing both the neural network model size and decoding graph size due to the increased number of modeling units. Finally, some methods add automatic speech recognition (ASR) as an auxiliary task during training.

\textbf{Pure neural models} abandon HMMs and completely rely on neural networks for acoustic modeling, where the subwords or even the whole word of the wake word phrase (wake phrase, for short) is directly used as modeling units. The first successful wake word detection systems of this type are proposed in \cite{chen2014small,sainath2015convolutional}. They use individual words in the wake phrase as modeling units to reduce the network size. However, they still need a forced alignment obtained from an existing HMM-based ASR system, to obtain training labels, which limits their applications if an ASR system is unavailable. For decoding, they adopt a fast posterior handling approach where the posterior of words is smoothed within a sliding window over the audio frames. \cite{myer2018efficient,coucke2019efficient} use the whole wake phrase as the training target, but it still needs phone-level alignments to pretrain a small network being part of a larger one. There are also several proposals that do not require frame-level alignment for training, including max-pooling \cite{sun2016max,hou2020mining}, the attention mechanism \cite{shan2018attention,wang2019adversarial}, and global mean-pooling \cite{bai2019time}.

It has been shown that a \emph{sequence-level training criteria} perform better than frame-level criteria for ASR.
The output in a wake word detection task, by contrast, is relatively simple. However, if the modeling units are subwords (e.g., phonemes or HMM states), wake word detection may still be considered as a sequence prediction task. Sequence-level discriminative training such as CTC loss \cite{graves2006connectionist} has been explored for the wake word detection task with graphemes or phonemes as subword units \cite{fernandez2007application,lengerich2016end,wang2017small,zhuang2016unrestricted}. Lattice-free maximum mutual information (LF-MMI) is an HMM-based sequence-level loss first proposed in \cite{povey2016purely} for ASR. In the context of wake word detection, it is recently investigated in \cite{chen2018sequence}, where it still requires alignments from a prior model like an HMM-GMM system to generate numerator graphs.

In this paper we propose a wake word detection system with alignment-free LF-MMI as training criterion, while not requiring any forced alignments for training.\footnote{The code and recipes are available in Kaldi~\cite{povey2011kaldi}:~\url{https://github.com/kaldi-asr/kaldi/tree/master/egs/{snips,mobvoi,mobvoihotwords}}.} Alignment-free LF-MMI was initially proposed for ASR \cite{hadian2018end}. In order to make it work for our task, we made several necessary adaptations/changes to the lexicon, HMM topology, data preprocessing for both efficiency and performance reasons. A fast online decoder is also proposed for our task. The experiments on three real wake word data sets all show its superior performance compared to other systems recently reported on the same data sets.

\section{The Proposed System}
\subsection{HMM Topology}
\begin{figure}[ht]
  \centering
  \begin{subfigure}{.4\textwidth}
  \centering
  \includegraphics[width=0.7\textwidth]{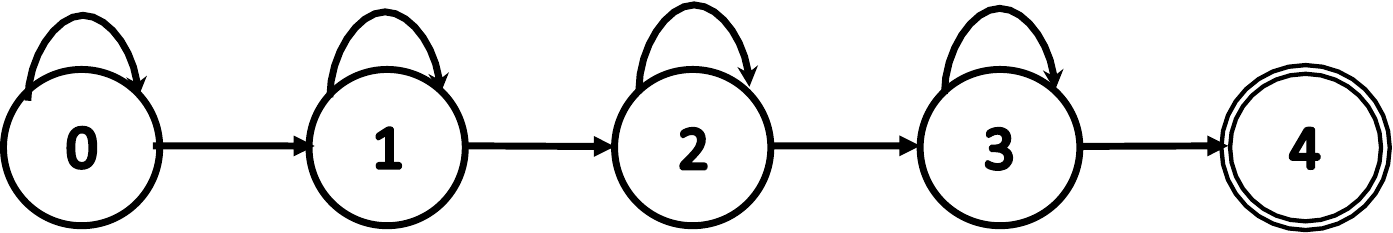}
  \end{subfigure}
  \begin{subfigure}{.45\textwidth}
  \centering
  \includegraphics[width=0.25\textwidth]{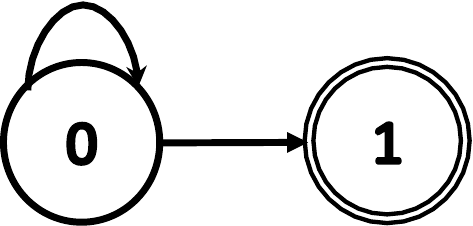}
  \end{subfigure}
  \caption{The HMM topologies used for the \emph{wake word} and \emph{freetext} (top), and \emph{SIL} (bottom). The number of emitting HMM states is 4 and 1, respectively. The final states are non-emitting.}
  \label{fig:hmm_topo}
  \vspace{-3mm}
\end{figure}

\vspace{-1mm}
\label{sec:hmm_topo}
Different from other traditional HMM-based keyword-filler models (e.g., those proposed in \cite{panchapagesan2016multi,sun2016compressed,wu2018monophone,chen2018sequence} where each phoneme of the whole wake phrase corresponds to an HMM or HMM state, we propose to model the whole wake phrase (in positive recordings) with a single HMM (referred to as \emph{word} HMM), and the number of distinct states within that HMM is a predefined value which is not necessarily proportional to the number of phonemes in its pronunciation. We argue that using a fixed number of HMM states, which is usually less than the number of the actual phonemes, has enough modeling power for the wake word task. Similarly, we use another HMM of the same topology (referred to as \emph{freetext} HMM) to model all non-silence speech (in negative recordings). From our preliminary experiments we also found that having an additional HMM dedicated to non-speech sounds, denoted \emph{SIL} and called the \emph{``silence'' phone}, is crucial for good performance. The \emph{SIL} phone is added as \emph{optional} silence \cite{chen2015pronunciation} to the beginning and end of each positive/negative recording so that it can learn the actual silence properly. The resulting topologies are shown in Fig.~\ref{fig:hmm_topo}.

\vspace{-2mm}
\subsection{Alignment-Free Lattice-Free MMI}
\vspace{-0.5mm}
\label{sec:lf_mmi}
Lattice-free MMI (LF-MMI) loss \cite{povey2016purely} is a sequence-level criterion and can be formulated as:
\begin{equation}
\mathcal{F}_{\textrm{LF-MMI}}=\sum_{n=1}^{N}\log P(L_n|\mathbf{O}_n)=\sum_{n=1}^{N}\log\frac{P(\mathbf{O}_n|L_n) P(L_n)}{\sum_L{P(\mathbf{O}_n|L)}P(L)} \nonumber
\end{equation}
where $L_n$ and $L$ are the subword truth sequence and a competing hypothesis sequence respectively, and $\mathbf{O}_n$ is the input audio. In the regular LF-MMI the numerator graph used to compute the truth sequence is an acyclic graph generated from an existing GMM model. In alignment-free LF-MMI \cite{hadian2018end}, the numerator graph is an unexpanded FST directly generated from training transcripts, giving more freedom to learn the alignments during the forward-backward pass in training.

\begin{figure}[ht]
  \centering
  \includegraphics[width=0.25\textwidth]{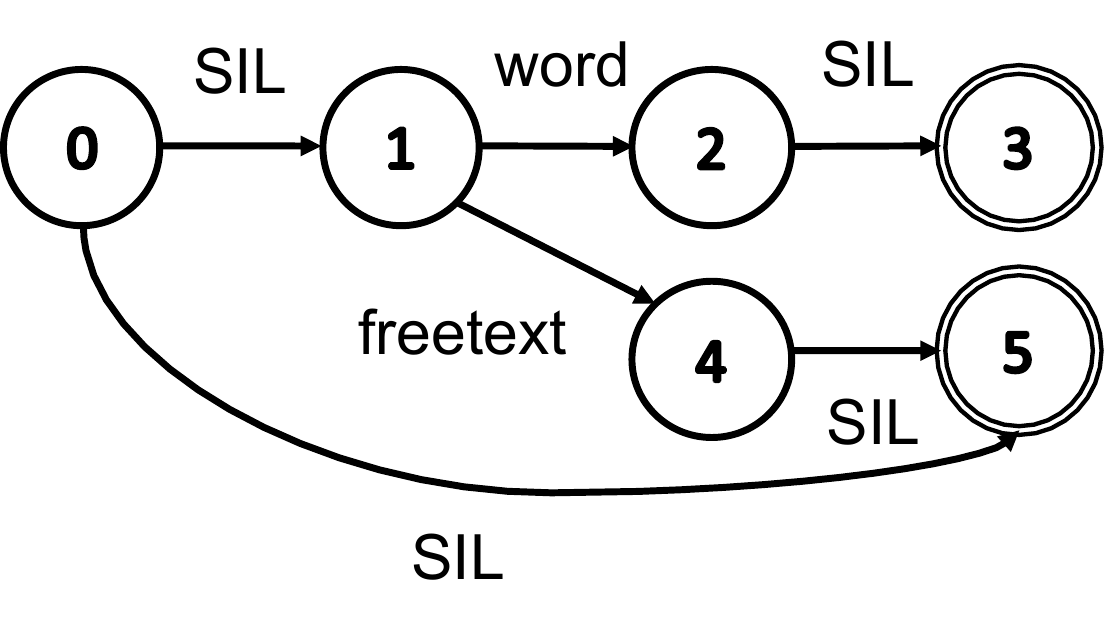}
  \caption{Topology of the phone language model FST for the denominator graph. Labels on arcs represent phones.}
  \label{fig:phonelm}
  \vspace{-4mm}
\end{figure}

For ASR, the competing hypotheses in the denominator graph are constructed from a phone LM trained from the training transcripts. On the contrary, for our wake word detection task, we manually specify the topology of the phone LM FST as in Fig.~\ref{fig:phonelm}. One path containing the \emph{word} HMM corresponds to positive recordings\footnote{If we have more than one wake word as those in our Mobvoi (SLR87) data set, each wake word would correspond to one such path.}, and the other two correspond to negative recordings (other speech/non-speech and silence). We assign final weights in a way such that they reflect the ratio of the number of positive/negative examples in the training set.

\vspace{-2mm}
\subsection{Acoustic Modeling}
\vspace{-0.5mm}
\label{sec:acoustic_modeling}
Owing to efficiency and latency concerns specific to our task, the family of recurrent \cite{hochreiter1997long,cho2014learning} or self-attention-based \cite{vaswani2017attention} neural networks is not within our consideration. Instead, we use factorized TDNN (TDNN-F) with skip connections \cite{povey2018semi} for acoustic modeling.
In a TDNN-F layer, the number of parameters is reduced by factorizing the weight matrix in TDNN layers \cite{peddinti2015time} into the product of two low-rank matrices, the first of which is constrained to be semi-orthogonal.
 
As in other architectures like ResNet \cite{he2016deep}, we incorporate skip connections: each TDNN-F layer receives its immediate prior layer’s output as the skip connection, which is added to the input of the current layer after being scaled down by 0.66.
 
We use a narrow (hidden dimension is 80) but deep (20 layers) network with each output frame covering a receptive field of size 80. The output is evaluated every 3 frames for LF-MMI loss to reduce the the computation cost both in training and test time. We also find a cross-entropy regularization together with the main LF-MMI loss helpful. As a result, the total number of parameters is about 150k, with the number of targets being only 18 using the HMM topologies described in Sec.~\ref{sec:hmm_topo}.

\vspace{-2mm}
\subsection{Data Preprocessing and Augmentation}
\vspace{-0.5mm}
\label{sec:data_preprocessing}
Compared with the positive recordings, the negative recordings usually have a longer duration and have more variability as they can include all possible speech except the wake phrase. However we only use a single \emph{freetext} HMM for it, making it difficult for the model to learn smoothly if batching them with positives directly (see Sec.~\ref{sec:exp_subseg}).
To tackle this problem, we chunk the negative recordings into shorter chunks of random lengths drawn from the empirical length distribution of the positive recordings, disregarding word boundaries. Successive chunks overlap by 0.3s, giving a trailing word-fragment from one chunk a chance to appear as a whole-word in the next. All chunks are assigned a negative label.

Although all our training data is recorded in real environments with background noise, we still found that data augmentation is helpful. Therefore we apply the same type of data augmentation techniques as used in \cite{wang2019jhu}, making use of noise, music, background speech from the MUSAN corpus \cite{snyder2015musan}, simulated reverberation \cite{ko2017study} and speed perturbation \cite{ko2015audio}. This effectively increases the amount of training data 7 times.

\vspace{-2mm}
\subsection{Decoding}
\vspace{-1mm}
\label{sec:decoding}
We next describe online Viterbi decoding without lattice generation for wake word detection, using the term ``tokens'' to denote partial hypotheses \cite{odell1994one}.

First we construct our decoding graph with a word-level FST specifying the prior probabilities of all possible word paths, in a similar way as we specify the phone language model FST in Fig.~\ref{fig:phonelm},  except that the start state and final states are merged to form a loop. The loop allows decoding with an audio interleaving with wake words and other possible speech.

\begin{algorithm}[ht]
\caption{Update the Immortal Token for Backtracking}
\label{code:update_immortal_token}
\hspace*{\algorithmicindent} \textbf{Input}: activeTokList \Comment{represents all current hypotheses} \\
\hspace*{\algorithmicindent} \textbf{Output}: immortalTok \Comment{is global, storing the latest one}
\begin{algorithmic}[1]
\Procedure{UpdateImmortalToken} {activeTokList}
\State emitting $\gets$ $\emptyset$
\For{tok in activeTokList} \label{code:update_immortal_token:init_begin}
    \While{isNonEmittingToken(tok)} tok $\gets$ tok.prev \EndWhile
    \If{tok $\neq$ NULL} emitting.insert(tok) \EndIf
\EndFor \label{code:update_immortal_token:init_end}
\State tokenOne $\gets$ NULL
\While{True} \label{code:update_immortal_token:track_begin}
    \If{$|$emitting$|$ = 1}
        \State tokenOne $\gets$ emitting[0]; break
    \EndIf
    \If{emitting = $\emptyset$} break \EndIf
    \State prevEmitting $\gets$ $\emptyset$
    \For{tok in emitting}
        \State prevTok $\gets$ tok.prev
        \While{isNonEmittingToken(tok)}
            \State prevTok $\gets$ tok.prev
        \EndWhile
        \If{prevTok = NULL} continue
        \EndIf
        \State prevEmitting.insert(prevTok)
    \EndFor
    \State emitting $\gets$ prevEmitting \label{code:update_immortal_token:track_end}
\EndWhile
\If{tokenOne $\neq$ NULL} \label{code:update_immortal_token:update_begin}
    \State immortalTok $\gets$ tokenOne \label{code:update_immortal_token:update_end}
\EndIf
\EndProcedure
\end{algorithmic}
\end{algorithm}
During online decoding, every time after processing a fixed-length chunk from a recording, we backtrack along the frames delimited by two most recent ``immortal tokens'' by calling the routine \textsc{UpdateImmortalToken} in Algorithm~\ref{code:update_immortal_token}, checking if there is a wake word detected from this partial backtracking. If a wake word is found, we just stop decoding and trigger the system; otherwise continue the decoding process. The ``immortal token'' is the common ancestor (prefix) of all active tokens, i.e. it will not ``die'' no matter which active token eventually survives.
Line~\ref{code:update_immortal_token:init_begin}-\ref{code:update_immortal_token:init_end} obtain the last emitting token from each active token. Line~\ref{code:update_immortal_token:track_begin}-\ref{code:update_immortal_token:track_end} are trying to find the common ancestor of all active tokens. Line~\ref{code:update_immortal_token:update_begin}-\ref{code:update_immortal_token:update_end} update the immortal token if a newer one is found; otherwise keep the old one from the previous decoding step.

The intuition is that, if all currently active partial hypotheses are from the same token at a previous time-step, all hypotheses before that token had already collapsed to one hypothesis (due to beam search pruning and token recombination), from which we would check whether it contains the wake word in a chunk-by-chunk fashion.

\vspace{-1mm}
\section{Experiments}
\label{sec:exp}
\vspace{-0.5mm}
\subsection{Data Sets}
\label{sec:exp_datasets}
\vspace{-1mm}
There are three real wake word data sets to be evaluated: SNIPS data set\footnote{\url{https://github.com/snipsco/keyword-spotting-research-datasets}} \cite{coucke2019efficient} with the wake word ``Hey Snips'', Mobvoi single wake word data set\footnote{This data set is not publicly available.} \cite{wang2019adversarial} with the wake word ``Hi Xiaowen'', and Mobvoi (SLR87) data set\footnote{~\url{https://www.openslr.org/87}} \cite{hou2019region} with two wake words ``Hi Xiaowen'' and ``Nihao Wenwen''. The statistics for each data set are summarized in Table~\ref{tab:datasets}. We will use the first two data sets to demonstrate the effects of several design choices in our system, and give the final results on all these three data sets when comparing our system with others. If not otherwise specified, we show our experimental results in an incremental way, meaning that later experiments would be conducted on top of the one that is better from the previous experiment. The operating points in DET curves are obtained by varying the cost corresponding to the positive path in the decoding graph while keeping the cost corresponding to negative path at 0. 40-dimensional MFCC features are extracted in all the experiments.

\begin{table}[htb]
  \caption{Statistics for the three wake word data sets.}
  \vspace{-5mm}
  \begin{center}
  \begin{adjustbox}{max width=\linewidth}
    \begin{tabular}{ c c c c c c c c c}
    \toprule
    \multirow{2}{*}{Name} & \multicolumn{2}{c}{Train} & \multicolumn{2}{c}{Dev} & \multicolumn{2}{c}{Eval} \\
    \cmidrule(lr){2-3} \cmidrule(lr){4-5} \cmidrule(lr){6-7}
    & \#Hrs & \#Utts (\#Positive) & \#Hrs & \#Utts (\#Positive) & \#Hrs & \#Utts (\#Positive) \\
    \midrule
    SNIPS & 54 & 50,658 (5,799) & 24 & 22,663 (2,484) & 25 & 23,072 (2,529) \\
    Mobvoi & 67 & 74,134 (19,684) & 7 & 7,849 (2,343) & 7 & 7,841 (1,942) \\
    Mobvoi (SLR87) & 144 & 174,592 (43,625\tablefootnote{~The statistics include two wake words.}) & 44 & 38,530 (7,357) & 74 & 73,459 (21,282) \\
    \bottomrule
    \end{tabular}
  \end{adjustbox}
  \end{center}
  \vspace{-5mm}
  \label{tab:datasets}
\end{table}

\vspace{-3mm}
\subsection{Effect of Negative Recordings Sub-segmentation}
\vspace{-1mm}
\label{sec:exp_subseg}
We first show the effect of sub-segmenting negative recordings on training. We start from the training data with only speed-perturbation applied. To keep consistent with the numbers reported in others' work on the same data sets, false rejection rate (FRR) is reported in Table~\ref{tab:subsegmentation} at 0.5 false alarms per hour (FAH) on the SNIPS data set, and at 1.5 FAH for Mobvoi. Apparently without sub-segmentation the performance is far from satisfactory, indicating the alignments learned with the LF-MMI system is poor. We also inspected the training/validation loss in both cases (not shown here), and found that there is severe overfitting when training without sub-segmentation.

\begin{table}[htb]
  \caption{Effect of sub-segmentation of negative recordings.}
  \vspace{-5mm}
  \begin{center}
    \begin{adjustbox}{max width=\linewidth}
    \begin{tabular}{ l c c}
    \toprule
    FRR(\%) & SNIPS (FAH=0.5) & Mobvoi (FAH=1.5) \\
    \midrule
       w/o sub-segmentation & 67 & 47 \\
       w/ sub-segmentation  &  \textbf{0.6}  & \textbf{5.6} \\
    \bottomrule
    \end{tabular}
    \end{adjustbox}
  \end{center}
  \label{tab:subsegmentation}
  \vspace{-3mm}
\end{table}

\vspace{-3.5mm}
\subsection{Effect of Data Augmentation}
\label{sec:exp_aug}
\vspace{-1mm}
We augment the training data using the MUSAN corpus in the following way: we randomly apply additive noise from the ``babble'', ``music'' and ``noise'' data sets separately on each copy of the
original training data once. For each training example, ``babble'' is added as background noises 3 to 7 times with SNRs ranging from 13 to 20; “music” is added as background noises once
with SNRs ranging from 5 to 15; “noise” is added as foreground
noises at the interval of 1 second with SNRs ranging from 0 to
15. Then reverberation is also separately applied to the training data using the simulated RIRs with room sizes uniformly sampled from 1 meter to 30 meters. The above procedure increase the training data by a factor of 4. We then apply 3 way speed-perturbation on top of the original training set. The augmentation strategy together results in about 7$\times$ more training data. The results before and after augmentation are shown in Table~\ref{tab:augmentation}. It can be seen the augmentation strategy is highly effective, where FRR with SNIPS is even 0 at FAH=0.5.

\begin{table}[htb]
  \caption{Effect of data augmentation.}
  \vspace{-5mm}
  \begin{center}
    \begin{adjustbox}{max width=\linewidth}
    \begin{tabular}{ l c c}
    \toprule
    FRR(\%) & SNIPS (FAH=0.5) & Mobvoi (FAH=1.5) \\
    \midrule
       w/o data augmentation & 0.6 & 5.6 \\
       w/ data augmentation  &  \textbf{0}  & \textbf{0.4} \\
    \bottomrule
    \end{tabular}
    \end{adjustbox}
  \end{center}
  \label{tab:augmentation}
  \vspace{-5mm}
\end{table}

\vspace{-3mm}
\subsection{Effect of Alignment-Free LF-MMI Loss}
\label{sec:exp_loss}
\vspace{-1mm}
To compare our proposed alignment-free LF-MMI loss with regular LF-MMI and conventional cross-entropy loss for our task, we train a phoneme-based HMM-GMM system to generate the numerator lattice (for regular LF-MMI loss) or the forced alignments (for conventional cross-entropy loss) for the same sub-segmented and augmented training data. the network architectures are the same as what is used for alignment-free LF-MMI training except the final layer (depending on the loss being used). The results in Table~\ref{tab:xent} validate that LF-MMI loss is generally advantageous to the frame-level cross-entropy loss in the wake word detection task, and alignment-free LF-MMI loss achieves better performance than regular LF-MMI on SNIPS and Mobvoi (SLR87), but worse on Mobvoi. It is worth noting that we believe SNIPS and Mobvoi (SLR87) results are more indicative of performance, as after manually listening to the false alarms in Mobvoi at this specific operating point, we found that all the false alarms (9 in total) are actually intentionally pronounced with a different tone on the last character ``wen'', which is extremely difficult for the model to learn given the limited amount of data; some would even argue that those examples should be labeled as positive cases in order for the model to accommodate Chinese speakers with accents. The better performance of the system with alignment-free LF-MMI loss is possibly due to the capability of learning flexible alignments than GMM models.

\begin{table}[htb]
  \caption{Effect of alignment-free LF-MMI loss.}
  \vspace{-5mm}
  \begin{center}
    \begin{adjustbox}{max width=\linewidth}
    \begin{tabular}{ l c c c c}
    \toprule
    \multirow{2}{*}{FRR(\%)} & SNIPS (FAH=0.5) & Mobvoi (FAH=1.5) & \multicolumn{2}{c}{Mobvoi (SLR87) (FAH=0.5)} \\
    \cmidrule(lr){4-5}
     & & & Hi Xiaowen & Nihao Wenwen \\
    \midrule
       cross-entropy & 0.6 & 3.5 & N/A & N/A \\
       regular LF-MMI & 0.1  & \textbf{0.2} & 0.6 & 0.7 \\
       alignment-free LF-MMI  &  \textbf{0}  & 0.4 & \textbf{0.4} & \textbf{0.5} \\
    \bottomrule
    \end{tabular}
    \end{adjustbox}
  \end{center}
  \label{tab:xent}
  \vspace{-3mm}
\end{table}

\vspace{-3mm}
\subsection{Regular LF-MMI Refinement}
\label{sec:exp_refine}
\vspace{-1mm}
The experiment from the previous section motivates us to do an additional experiment investigating how the regular LF-MMI performs when it gets alignments from the alignment-free LF-MMI system instead of from a GMM model, and whether the regular LF-MMI training could further improve the performance as a refinement of our existing system. To this end, we compare the three systems in Table~\ref{tab:refine}. Note that as we already achieve FRR=0 at FAH=0.5 with SNIPS using our alignment-free LF-MMI system, we set the operating point at a smaller FAH (0.04) for it. Table~\ref{tab:refine} demonstrates that further improvement can be obtained by running an additional regular LF-MMI training on top of alignment-free LF-MMI, suggesting an optional refinement stage for better performance.

\begin{table}[htb]
  \caption{Effect of using alignments from Alignment-free LF-MMI for regular LF-MMI.}
  \vspace{-5mm}
  \begin{center}
    \begin{adjustbox}{max width=\linewidth}
    \begin{tabular}{ l c c}
    \toprule
    FRR(\%) & SNIPS (FAH=0.04) & Mobvoi (FAH=1.5) \\
    \midrule
       regular LF-MMI & 0.2  & \textbf{0.2} \\
       alignment-free LF-MMI  & 0.2 & 0.4 \\
       \quad+regular LF-MMI refinement & \textbf{0.1} & 0.3 \\
    \bottomrule
    \end{tabular}
    \end{adjustbox}
  \end{center}
  \label{tab:refine}
  \vspace{-5mm}
\end{table}

\vspace{-3mm}
\subsection{Comparison with Other Baseline Systems}
\begin{figure}[ht]
  \centering
  \includegraphics[width=0.5\textwidth]{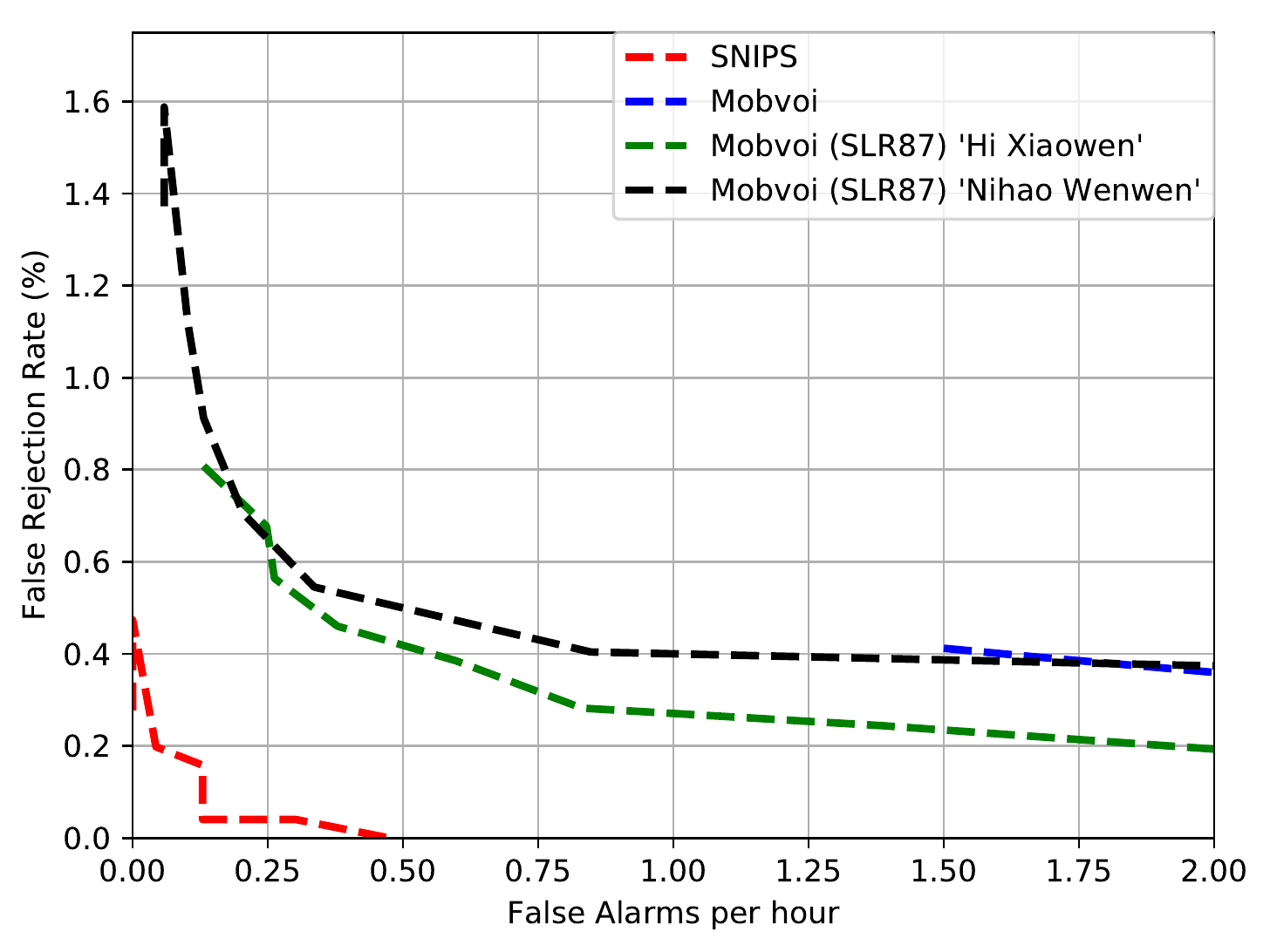}
  \caption{DET curves for the three data sets.}
  \label{fig:det_all}
  \vspace{-3mm}
\end{figure}

\vspace{-1mm}
\label{sec:exp_comp}
We compare our proposed system with other systems recently proposed on the same data sets. We use our alignment-free LF-MMI system without refinement as the refinement is optional. The results are shown in Table~\ref{tab:comparisons}. DET curves of our system on all the three data sets are plotted in Fig.~\ref{fig:det_all}.

For SNIPS data set we compare against their original paper \cite{coucke2019efficient}, where a voice activity detection system is used to obtain frame-level wake word labels for training. While their system is already very good in term of FRR, our system even achieves FRR=0 at FAH=0.5.

For Mobvoi data set we compare our system with \cite{wang2019adversarial} where an attention mechanism is adopted for pooling across frames to make a prediction. They also propose an adversarial examples generation algorithm for robust training. The modeling units are wake words, and they use recurrent rather than convolutional networks. Our system achieves significant better results, improving FRR by around 90\% relative compared to their number at FAH=1.5.

For Mobvoi (SLR87) data set, The approach proposed in \cite{hou2020mining} are compared, where the label imbalance issue is tackled by selective negative sampling. Again, it uses wake words as modeling units. Note that this data set contains two wake words (``Hi Xiaowen'' and ``Nihao Wenwen''), and when we are evaluating for a specific one, the other one is considered as negative. We achieve 50-70\% reduction in FRR than the baseline with the same FAH=0.5.

\begin{table}[htb]
  \caption{Comparison with other wake-word detection baselines.}
  \vspace{-5.5mm}
  \begin{center}
    \begin{adjustbox}{max width=\linewidth}
    \begin{tabular}{ l c c c}
    \toprule
    \textit{SNIPS} & \#Params & \multicolumn{2}{c}{FRR(\%) at FAH=0.5} \\
      Coucke at al. \cite{coucke2019efficient} & 220k & \multicolumn{2}{c}{0.12} \\ 
      alignment-free LF-MMI (Ours) & 150k & \multicolumn{2}{c}{\textbf{0}} \\
    \midrule
    \textit{Mobvoi} & \#Params & \multicolumn{2}{c}{FRR(\%) at FAH=1.5} \\
      Wang at al. \cite{wang2019adversarial} & 84k & \multicolumn{2}{c}{$\sim$3.6} \\ 
      alignment-free LF-MMI (Ours) & 150k & \multicolumn{2}{c}{\textbf{0.4}} \\
    \midrule
    \multirow{2}{*}{\textit{Mobvoi (SLR87)}} & \multirow{2}{*}{\#Params} & \multicolumn{2}{c}{FRR(\%) at FAH=0.5} \\
    \cmidrule(lr){3-4}
     & & Hi Xiaowen & Nihao Wenwen \\
       Hou at al. \cite{hou2020mining}~\tablefootnote{~The numbers shown here, different from those in the original paper, are obtained at \url{https://github.com/jingyonghou/KWS_Max-pooling_RHE} on the same data as what we are using.} & N/A & 1.3 & 1.0 \\
       alignment-free LF-MMI (Ours) & 150k &  \textbf{0.4} &  \textbf{0.5} \\
    \bottomrule
    \end{tabular}
    \end{adjustbox}
  \end{center}
  \label{tab:comparisons}
  \vspace{-6mm}
\end{table}

\vspace{-4mm}


\vspace{-3mm}
\section{Conclusions and Future Work}
\vspace{-1mm}
We describe a suite of methods to build a hybrid HMM-DNN system for wake word detection, including sequence-discriminative training based on \emph{alignment-free} LF-MMI loss, removing the need for frame-level training alignments, and whole-word HMMs for the wake word and filler speech, removing the need for training transcripts or pronunciation lexicons.  These features significantly reduce model sizes and greatly simplify the training process. An online decoder tailored to wake word detection is proposed to complete the suite. The system widely outperforms other wake word detection systems on three different real-world wake word data sets. We have open-sourced our system in Kaldi, and to make it accessible to other deep learning frameworks, we are implementing a PyTorch-based version based on \espresso \cite{wang2019espresso} and \pychain \cite{shao2020pychain}.

\bibliographystyle{IEEEtran}

\bibliography{mybib}

\end{document}